 \documentclass[12pt]{article}
 \usepackage{amsfonts}
 \usepackage{amsmath}
 \usepackage{amscd}
 \usepackage{amsthm}
 \def\R{\mathbb{R}}
 
 \def\Z{\mathbb{Z}}

\usepackage{latexsym}
 \numberwithin{equation}{section}
 \newtheorem{thm}{Theorem}[section]
 \newtheorem{lem}{Lemma}[section]
 \newtheorem{prop}{Proposition}[section]
 \theoremstyle{definition}

\title{ Stability of periodic
traveling waves for complex modified Korteweg-de Vries equation}
\author{Sevdzhan Hakkaev$ ^1$, Iliya D. Iliev$ ^2$  and
   Kiril Kirchev$ ^2$ }

\begin{document}

\maketitle

\begin{center}
{$ ^1$Faculty of Mathematics and Informatics,

 Shumen University, 9712 Shumen, Bulgaria

 \vspace{2ex}
 $ ^2$Institute of Mathematics and Informatics,

 Bulgarian Academy of Sciences, 1113 Sofia, Bulgaria \\ }
 \end{center}

\begin{abstract} We study the existence and stability of
periodic traveling-wave solutions for complex modified
Korteweg-de Vries equation. We also discuss the problem of uniform
continuity of the data-solution mapping.
\end{abstract}

\section{Introduction}

 Consider the complex modified Korteweg-de Vries equation
   \begin{equation}\label{1.1}
     u_{t}+6|u|^{2}u_{x}+u_{xxx}=0,
   \end{equation}
  where $u$ is a complex-valued function of $(x,t)\in\R^2$.
  In this paper, we study the orbital stability of the family of periodic
  traveling-wave solutions
\begin{equation}\label{1.2}
u=\varphi(x,t)=e^{i\omega (x+(3a+\omega^2)t)}r(x+(a+3\omega^2) t).
\end{equation}
where $r(y)$ is a real-valued $T$-periodic function and $a,
\omega\in\R$ are parameters. The problem of the stability of
solitary waves for nonlinear dispersive equations goes back to the
works of Benjamin \cite{Be1} and Bona \cite{Bo} (see also
\cite{AlBoHe, W1, W2}). A general approach for investigating the
stability of solitary waves for nonlinear equations having a group
of symmetries was proposed in \cite{GSS}. The existence and
stability of solitary wave solutions for equation (\ref{1.1}) has
been studied in \cite{ZIK}. In contrast to solitary waves for
which stability is well understood, the stability of periodic
traveling waves has received little attention. Recently in \cite{ABS},
the authors developed a complete theory on the stability of cnoidal waves for
the KdV equation. Other new explicit formulae for the periodic traveling
waves based on the Jacobi elliptic functions, together with their
stability, have been obtained in \cite{An2, AnNa, HIK} for the
nonlinear Schr\"odinger equation, modified KdV equation, and
generalized BBM equation. In \cite{Ha}, the stability of periodic traveling
wave solutions of BBM equation which wave profile stays close to the constant
state $u=(c-1)^{1/p}$ is considered.

Our purpose here is to study existence and stability of periodic
traveling wave solutions of equation (\ref{1.1}).
 We base our analysis on the invariants $Q(u), P(u),$ and $F(u)$
(see Section 4). Our approach is to verify that $\varphi$ is a minimizer
of a properly chosen functional $M(u)$ which is conservative with respect
to time over the solutions of (\ref{1.1}). We consider the $L^2$-space of
$T$-periodic functions in $x\in\R$, with a norm $||.||$ and a scalar
product $\langle .,.\rangle$. To establish that the orbit
$${\cal O}=\{e^{i\omega\eta}\varphi(\cdot-\xi,t):\;
\omega\in 2\pi\Z/T,\; \xi,\eta\in[0,T]\}$$
is stable, we take $u(x,t)=e^{i\omega \eta}\varphi(x-\xi,t)+h(x,t)$,
$h=h_1+ih_2$ and express the leading term of $M(u)-M(\varphi)$ as
$\langle L_1h_1,h_1\rangle+\langle L_2h_2,h_2\rangle$ where
$L_i$ are second-order selfadjoint differential operators in $L^2[0,T]$
with potentials depending on $r$ and satisfying $L_1 r'=L_2r=0$. The proof of
orbital stability requires that zero is the second eigenvalue of $L_1$ and
the first one of $L_2$. Therefore, we are able to establish stability when
$r(y)$ does not oscillate around zero. Sometimes, the waves (\ref{1.2})
with this property are called "dnoidal waves" because $r$ is expressed by
means of the elliptic function $dn(y;k)$.

The paper is organized as follows. In Section 2, we discuss in brief the
correctness of the Cauchy problem for (\ref{1.1}) in periodic Sobolev
spaces $H^s$, $s\in\R$ (equipped with a norm $||.||_s$). The problem is
locally well-posed for $s>\frac32$ and ill-posed for $s<\frac12$. In
Section 3 we outline the existence and the properties of the periodic
traveling-wave solutions (\ref{1.2}) to (\ref{1.1}), with emphasis on
the case when $r$ does not oscillate around zero. In Section 4 we prove
our main orbital stability result (Theorem \ref{t31}). In the Appendix, we
establish some technical results we need during the proof of our main
theorem.

%%%%%%%%%%%%%%%%%%%%%%%%%%%%%%%%%%%%%%%%%%%%%%%%%%%%%%%%%%%%%%%%%%%%%%%%%

%%%%%%%%%%%%%%%%%%%%%%%%%%%%%%%%%%%%%%%%%%%%%%%%%%%%%%%%%%%%%%%%%%%%%%%%%

 \section{Cauchy problem}

  In this section we discuss the well-posedness of the initial-value problem
  for the complex modified Korteweg-de Vries equation in the
  periodic case. We take an initial value $u_0(x)$ in a periodic Sobolev
  space $H^s=H^s[0,T]$. Local well-posedness means that there exists a
  unique solution $u(.,t)$ of (\ref{1.1}) taking values in $H^s$ for a time
  interval $[0, t_0)$, it defines a continuous curve in $H^s$ and
  depends continuously on the initial data.

     The local well-posedness for (\ref{1.1}) in the non-periodic case is
     studied in \cite{ZIK} by applying Kato's theory of abstract quasilinear
     equations \cite{Ka1, Ka2}. In the periodic case the conditions are
     verified in the same way as in \cite{ZIK}, therefore we present
     here without proving the following result.

  \begin{thm}\label{ts1}
    Let $s>\frac{3}{2}$ and $T>0$. For each $u_0(x)\in H^s[0,T]$
    there exists
    $t_0$ depending only on $||u_{0}||_s$ such that $(\ref{1.1})$
    has a unique solution $u(x,t)$, with $u(x,0)=u_0(x)$ and
      $$u\in C([0,t_0); H^s)\cap C^{1}([0, t_0); H^{s-3}).$$
      Moreover, the mapping $u_0(x)\rightarrow u(x,t)$ is
      continuous in the $H^s[0,T]$-norm.
   \end{thm}

  Sometimes it is more appropriate to consider other version of
  well-posedness, for example by strengthening our definition,
 requiring that the mapping data-solution is uniformly continuous, that is:
     for any $\varepsilon$, there exists $\delta>0$, such that if
     $||u_{01}-u_{02}||_s<\delta$, then $||u_{1}-u_{2}||_s<\varepsilon$, with
     $\delta=\delta(\varepsilon, M)$, where $||u_{01}||_s\leq
     M$ and $||u_{02}||_s\leq M.$ The ill-posedness of some
     classical nonlinear dispersive equations (KdV, mKdV, NLS) in
     both periodic and non-periodic cases are studied in \cite{BKPS,
     BPS, BGT, KPV1}. The approach in these papers is based on the existence
and good properties of the traveling wave solutions associated to the
respective equations.
   Below we discuss the problem of the uniform-continuity of
   data-solution mapping for (\ref{1.1}) in periodic Sobolev spaces
   with small $s$.

  \begin{thm}\label{ts11}
    The initial value problem for the complex modified Korteweg-de Vries
    equation $(\ref{1.1})$ is locally ill-posed for initial data in the
    periodic spaces $H^s$ with $s<\frac12$.
  \end{thm}
    {\bf Proof.} It is easy to see that
      \begin{equation}\label{4.2}
         u_{N,A}(x,t)=A\exp(i(Nx+(N^{3}-6A^2N)t))
      \end{equation}
    where $A$ is a real constant and $N$ is a positive integer, solves the
    equation (\ref{1.1}) with initial data $u_{0}(x)=A\exp(iNx)$.
    For $A=\alpha N^{-s}$ where $\alpha$ is a real parameter, we have
      $$\begin{array}{ll}
          ||u_{0}||_{s} \leq C\alpha^{2} \\
          \\
          ||u_{N,A}(\cdot ,t)||_{s}\leq C\alpha^{2}
       \end{array}
      $$
   with $C>0$. Let $A_{1}=\alpha_{1}N^{-s}$ and $A_{2}=\alpha_{2} N^{-s}$.
   For the Sobolev norm of the difference of two initial data, we have
     $$\begin{array}{ll}
         ||u_{A_{1}, N}(0)-u_{A_{2}, N}(0)||_{s}^{2}& =
\sum_{\xi\in\Z}(1+\xi^{2})^{s}|\widehat{u_{A_{1},N}}(\xi)-
\widehat{u_{A_{2},N}}(\xi)|^{2} \\
         \\
         & \leq C|\alpha_{1}-\alpha_{2}|^{2} \rightarrow 0 \;
         \; \rm{as} \; \; \alpha_{1} \rightarrow \alpha_{2}
      \end{array}
    $$
  On the other hand, we have
    $$\begin{array}{cc}
        ||u_{A_{1}, N}(\cdot ,t)-u_{A_{2}, N}(\cdot ,t)||_{s}^{2}\\
        \\  =\sum_{\xi\in\Z}(1+\xi^{2})^{s}|\widehat{u_{A_{1},N}}(\xi)
        -\widehat{u_{A_{2},N}}(\xi)|^{2} \\
        \\ =(1+N^{2})^{s}|\alpha_{1}N^{-s}\exp(i(N^{3}
        -6\alpha_{1}^{2}N^{-2s+1}))-\alpha_{2}N^{-s}
        \exp(i(N^{3}-6\alpha_{2}^{2}N^{-2s+1}))|^{2} \\
        \\
        \geq C|\alpha_{1}-\alpha_{2}
        \exp(i6(\alpha_{1}^{2}-\alpha_{2}^{2})N^{1-2s})|^{2}
      \end{array}
    $$
  Let $s<{\frac{1}{2}}$, and $\alpha_{1}$, $\alpha_{2}$ be chosen so that
     $$(\alpha_{1}^{2}-\alpha_{2}^{2})N^{1-2s}=C N^{2\nu},$$
    where $\nu >0$ and $2\nu+2s-1<0$. Then for
    $t={\frac{\pi}{2}}C^{-1}N^{-2\nu}$, we have
      $$||u_{A_{1}, N}(x,t)-u_{A_{2}, N}(x,t)||_{s}^{2}
      \geq C(\alpha_{1}^{2}+\alpha_{2}^{2})$$
   Note that $t$ can be made arbitrary small by choosing $N$ sufficiently large.
   This completes the proof of the theorem.  $\Box$

%%%%%%%%%%%%%%%%%%%%%%%%%%%%%%%%%%%%%%%%%%%%%%%%%%%%%%%%%%%%%%%%%%%%%%%%%%

%%%%%%%%%%%%%%%%%%%%%%%%%%%%%%%%%%%%%%%%%%%%%%%%%%%%%%%%%%%%%%%%%%%%%%%%%%
  \section{Periodic traveling-wave solutions}
We are looking for traveling-wave solutions for equation (\ref{1.1})
in the form
    \begin{equation}\label{2.1}
      \varphi (x,t)=e^{i\omega (x+\alpha t)}r(x+\beta t)
    \end{equation}
 where $\alpha,\beta,\omega\in\R$ and
 $r(y)$ is a smooth real periodic function with a given period $T$.
Substituting (\ref{2.1}) into (\ref{1.1})
and separating real and imaginary parts,
we obtain the following equations
\begin{equation}\label{2.3}
\begin{array}{l}
\beta-3\omega^2=\frac13(\alpha-\omega^2)=a\in\R,\\[2mm]
r''+2r^3+ar=0.\end{array}
\end{equation}
Therefore
\begin{equation}\label{1.3}
u=\varphi(x,t)=e^{i\omega (x+(3a+\omega^2)t)}r(x+(a+3\omega^2) t).
\end{equation}
Integrating once again the second equation in (\ref{2.3}), we obtain
\begin{equation}\label{newton}
r'^2=c-ar^2-r^4,
\end{equation}
hence the periodic solutions are given by the periodic trajectories
$H(r,r')=c$ of the Hamiltonian vector field $dH=0$ where
$$H(x,y)=y^2+x^4+ax^2.$$
Clearly, two cases appear:

\vspace{1ex}
\noindent
1) {\it Global center} $(a\geq 0)$. Then for any $c>0$ the orbit defined by
$H(r,r')=c$ is periodic and oscillates around the center at the origin.

\vspace{1ex}
\noindent
2) {\it Duffing oscillator} $(a<0)$. Then there are two possibilities

\vspace{1ex}
\noindent
2.1) ({\it outer case}): for any $c>0$ the orbit defined by $H(r,r')=c$
is periodic and oscillates around the eight-shaped loop $H(r,r')=0$ through
the saddle at the origin.

\vspace{1ex}
\noindent
2.2) ({\it left and right cases}): for any $c\in(-\frac14a^2,0)$ there are two
periodic orbits defined by $H(r,r')=c$ (the left and right ones). These are
located inside the eight-shaped loop and oscillate around the centers at
$(\mp\sqrt{-a/2},0)$, respectively.

\vspace{1ex}
In cases 1) and 2.1) above, $r(x)$ oscillates around zero and for this reason
we are unable to study stability properties of the wave (\ref{1.3}).
In the rest of the paper, we will consider the left and right cases of
Duffing oscillator.

\vspace{2ex}
\noindent
{\bf Remark.} One could also consider equation (\ref{1.1}) with a minus sign,
    $$ u_{t}-6|u|^{2}u_{x}+u_{xxx}=0.$$
It has a traveling-wave solution of the form (\ref{1.3}) where $r$
is a real-valued periodic function of period $T$ satisfying equation
$ r''-2r^3+ar=0.$ Taking $H(x,y)=y^2-x^4+ax^2,\quad a>0$
({\it the Truncated pendulum Hamiltonian}), we see that for
$c\in (0, \frac14a^2)$ the periodic solutions are given by the periodic
trajectories $H(r,r')=c$ of the Hamiltonian vector field $dH=0$ which
oscillate around the center at the origin and are bounded by the separatrix
contour $H(r,r')=\frac14a^2$ connecting the saddles $(\mp\sqrt{a/2},0)$.
Therefore, we are unable to handle this case, too.

\vspace{2ex}
In the left and the right cases, let us denote by $r_0>r_1>0$ the positive
roots of $r^4+ar^2-c=0$. Then, up to a translation, we obtain the respective
explicit formulas
\begin{equation}\label{1.7}
r(z)=\mp r_0 dn(\alpha z; k),\quad k^2=\frac{r_0^2-r_1^2}{r_0^2}
=\frac{a+2r_0^2}{r_0^2},
\quad \alpha=r_0, \quad T=\frac{2K(k)}{\alpha}.
\end{equation}
Here and below $K(k)$ and $E(k)$ are, as usual, the complete elliptic
integrals of the first and second kind in a Legendre form.
By (\ref{1.7}), one also obtains $a=(k^2-2)\alpha^2$ and, finally,
\begin{equation}\label{1.8}
T=\frac{2\sqrt{2-k^2}K(k)}{\sqrt{-a}}, \quad
k\in(0,1), \quad T\in I=\left(\frac{2\pi}{\sqrt{-2a}},\infty\right).
\end{equation}

\vspace{2ex} \noindent {\bf Lemma 3.1.} {\it For any
$a<0$ and $T\in I$, there is a constant $c=c(a)$
such that the periodic traveling-wave solution $(\ref{1.3})$ determined
by $H(r,r')=c(a)$ has a period $T$. The function $c(a)$ is
differentiable.}

\vspace{2ex}
\noindent
{\bf Proof.} The statement follows from the implicit function theorem.
It is easily seen that the period $T$ is a strictly increasing function of $k$:
$$\frac{d}{dk}(\sqrt{2-k^2}K(k))=\frac{(2-k^2)K'-kK}{\sqrt{2-k^2}}=
\frac{K'+E'}{\sqrt{2-k^2}}>0.$$
Given $a$ and $c$ in their range, consider the functions $r_0(a,c)$,
$k(a,c)$ and $T(a,c)$ given by the formulas we derived above.
We obtain
$$\frac{\partial T}{\partial c}=\frac{dT}{dk} \frac{dk}{dc}=
\frac{1}{2k}\frac{dT}{dk}\frac{d(k^2)}{dc}.$$
Further, we have in the left and right cases
$$\frac{d(k^2)}{dc}=\frac{d(k^2)}{d(r_0^2)} \frac{dr_0^2}{dc}
=-\frac{a}{r_0^4(a+2r_0^2)}.$$
We see that $\partial T(a,c)/\partial c\neq 0$, therefore
the implicit function theorem yields the result. $\Box$

%%%%%%%%%%%%%%%%%%%%%%%%%%%%%%%%%%%%%%%%%%%%%%%%%%%%%%%%%%%%%%%%%%%%%%%%%%
  \section{Stability}
In this section we prove our main stability result which concerns the
left (right) Duffing oscillator cases. Take $a<0$, $T>2\pi/\sqrt{-2a}$
and determine $c=c(a)$ so that the two orbits given by $H(r,r')=c$
have period $T$. Next, chose $\omega\neq 0$ in (\ref{1.3}) to satisfy
$\omega T/2\pi\in\Z$. Then $\varphi(x,t)$ is a solution of (\ref{1.1})
having a period $T$ with respect to $x$.

    \vspace{2ex}
   {\bf 1. Basic statements and reductions}
%%%%%%%%%%%%%%%%%%%%%%%%%%%%%%%%%%%%%%%%%%%%%%%%%%%%%%%%%%%%%%%%%%%

Take a solution $u(x,t)$ of (\ref{1.1}) of period $T$ in $x$ and
introduce the pseudometric
    \begin{equation}\label{3.1}
      d(u, \varphi)=\inf_{(\eta, \xi)\in \R^2}
      ||u(x,t)-e^{i\omega \eta}\varphi(x-\xi, t)||_{1}.
    \end{equation}

   The equation (\ref{1.1}) possesses the following conservation laws
     $$ Q(u)=i\int_{0}^{T}{\overline{u}_{x}u}dx,\quad
          P(u)=\int_{0}^{T}{|u|^{2}}dx,\quad
          F(u)=\int_{0}^{T}{(|u_{x}|^{2}-|u|^{4})}dx.$$
  Let
    $$M(u)=F(u)+(\omega^{2}-a)P(u)-2\omega Q(u).$$
  For a fixed $q>0$, we denote
       \begin{equation}\label{3.2}
         \begin{array}{ll}
          d_{q}^{2}(u,\varphi)&=\inf_{(\eta, \xi)\in \R^2}
 \left(||u_{x}(x,t)-e^{i\omega \eta}\varphi_{x}(x-\xi, t)||^2 \right.\\
&\left. +q||u(x,t)-e^{i\omega \eta}\varphi(x-\xi, t)||^2\right).
         \end{array}
        \end{equation}

Clearly, the infimum in  $(\ref{3.1})$ and $(\ref{3.2})$ is attained at some
point $(\eta, \xi)$ in the square $[0,T]\times [0,T]$. Moreover, for
$q\in [q_1,q_2]\subset(0,\infty)$, (\ref{3.2}) is a pseudometric equivalent
to (\ref{3.1}).
Now, we can formulate our main result in the paper.
       \begin{thm}\label{t31}   Let $\varphi$ be given by $(\ref{1.2})$, with $r\neq 0$.
         For each $\varepsilon>0$ there exists $\delta>0$ such that if
         $u(x,t)$ is a solution of $(\ref{1.1})$ and
         $d(u, \varphi )_{|t=0}<\delta$, then $d(u, \varphi)<\varepsilon$
         $\forall t\in [0,\infty)$.
       \end{thm}

The crucial step in the proof will be to verify the following statement.
      \begin{prop}\label{p31}
        There exist positive constants $m, q, \delta_{0}$ such that if $u$
        is a solution of $(\ref{1.1})$ such that $P(u)=P(\varphi)$ and
        $d_{q}(u,\varphi)<\delta_{0}$, then
          \begin{equation}\label{3.3}
            M(u)-M(\varphi)\geq md_{q}^{2}(u,\varphi).
          \end{equation}
      \end{prop}
 The proof consists of several steps. The first one concerns the metric
 $d_q$ introduced above.
      \begin{lem}\label{l31}
The metric $d_{q}(u,\varphi)$ is a continuous function of $t\in
[0, \infty)$.
      \end{lem}

\noindent {\bf Proof.} The proof of the lemma is similar to the proof of
Lemmas 1, 2 in \cite{Bo} $\Box$.

\vspace{1ex}
\noindent
      We fix $t\in [0,\infty)$ and assume that the minimum in (\ref{3.1})
      is attained at the point $(\eta, \xi)=(\eta(t), \xi(t))$. In order
      to estimate $\Delta M =M(u)-M(\varphi)$, we set
        $$u(x,t)=e^{i\omega \eta}\varphi(x-\xi, t)+h(x,t)$$
     and integrating by parts in the terms containing $h_{x}$ and
     $\overline{h}_{x}$, we obtain
      $$ \begin{array}{ll}
         \Delta M &=M(u)-M(\varphi) \\[1mm]
         &=2Re\int_0^T e^{i\omega \eta}\left[ -\varphi_{xx}+
         (\omega^2-a-2|\varphi |^2)\varphi +2i\omega \varphi_x\right]
         \overline{h}dx\\[1mm]
         &+\int_0^T\left[|h_{x}|^2+(\omega^2-a-4|\varphi|^2)|h|^2
         -2i\omega h\overline{h}_x-2Re\left(e^{-2i\omega \eta}
         \overline{\varphi}^2h^2\right)\right]dx\\[1mm]
         &-\int_0^T |h|^2\left(4Re\left(e^{i\omega \eta}\varphi
         \overline{h}\right)+|h|^2\right)dx\\[1mm]
         &=I_{1}+I_{2}+I_{3}.
      \end{array}
      $$
      Note that the boundary terms annihilate by periodicity.
 Using that $r(x)$ satisfies equation (\ref{2.3}), we obtain $I_1=0$. Let
      $$h=(h_{1}+ih_{2})e^{i\omega(x-\xi+(\omega^{2}+3a)t+\eta)}, $$
    where $h_{1}$ and $h_{2}$ are real periodic functions with period $T$.
    Then we have
       $$ \begin{array}{ll}
            |h|^{2}=h_{1}^{2}+h_{2}^{2} \\[1mm]
      |h_{x}|^{2}=(h_{1x}-\omega h_{2})^{2}+(h_{2x}+\omega h_{1})^{2}\\[1mm]
    \int_{0}^{T}{\overline{h}_{x}h}dx=i\int_{0}^{T}{[h_{2}h_{1x}-h_{1}h_{2x}
   -\omega (h_{1}^{2}+h_{2}^{2})]}dx\\[1mm]
            Re(h^{2}\overline{\varphi}^{2}e^{-2i\omega \eta})
            =r^{2}(h_{1}^{2}-h_{2}^{2}) \end{array}
       $$
       Thus for $I_2$ we obtain the expression
         $$ \begin{array}{ll}
              I_{2} &=\int_{0}^{T}{[h_{1x}^{2}-(a+6r^{2})h_{1}^{2}]}dx+
              \int_{0}^{T}{[h_{2x}^{2}-(a+2r^{2})h_{2}^{2}]}dx \\[1mm]
              &=M_{1}+M_{2}
            \end{array}
         $$
       Introduce in $L^2[0,T]$ the self-adjoint operators $L_1$ and $L_2$
       generated by the differential expressions
        $$\begin{array}{ll}
           L_{1}=-\partial_{x}^{2}-(a+6r^{2}), \\[2mm]
           L_{2}=-\partial_{x}^{2}-(a+2r^{2}),
          \end{array} $$
         with periodic boundary conditions in $[0,T]$.

        \vspace{2ex}
       {\bf 2. Spectral analysis of the operators $L_{1}$ and $L_{2}$}
%%%%%%%%%%%%%%%%%%%%%%%%%%%%%%%%%%%%%%%%%%%%%%%%%%%%%%%%%%%%%%%%%%%%%%%%%%

     Consider in $L^2[0,T]$ the following periodic eigenvalue problems
     \begin{equation}\label{3.4}
         \left\{
          \begin{array}{ll}
          L_1\psi=\lambda \psi\quad \mbox{\rm in}\;\,[0,T], \\
          \psi(0)=\psi(T), \; \; \psi'(0)=\psi'(T),\\
        \end{array} \right.
     \end{equation}
     \begin{equation}\label{3.5}
      \left\{
        \begin{array}{ll}
          L_2\chi=\lambda \chi \quad \mbox{\rm in}\;\,[0,T],  \\
          \chi(0)=\chi(T), \; \; \chi'(0)=\chi'(T).\\
        \end{array} \right.
    \end{equation}
    The problems (\ref{3.4}) and (\ref{3.5}) have each a countable infinite
    set of eigenvalues $\{ \lambda_{n} \}$ with $\lambda_n \rightarrow \infty$.
    We shall denote by $\psi_n$, respectively by $\chi_n$, the eigenfunction
    associated to the eigenvalue $\lambda_n$. For the periodic eigenvalue
    problems (\ref{3.4}) and (\ref{3.5}) there are associated semi-periodic
    eigenvalue problems in $[0,T]$, namely (e.g. for (\ref{3.5}))
    \begin{equation}\label{3.6}
      \left\{
        \begin{array}{ll}
          L_2\vartheta=\mu\vartheta \quad \mbox{\rm in}\;\,[0,T],\\
          \vartheta(0)=-\vartheta(T), \; \; \vartheta'(0)=-\vartheta'(T).\\
        \end{array} \right.
    \end{equation}
  As in the periodic case, there is a countable infinity set of eigenvalues
  $\{\mu_n\}$. Denote by $\vartheta_n$ the eigenfunction associated to the
  eigenvalue $\mu_n$. From the Oscillation Theorem \cite{MaWi} we know that
     $\lambda_0< \mu_0 \leq \mu_1<\lambda_1 \leq \lambda_2 <\mu_2 \leq \mu_3,
     \ldots$, $\lambda_0$ is simple and
     $$ \begin{array}{ll}
     \rm{(a)} \;\;\chi_0\;\;\mbox{\rm has no zeros on}\;\;[0,T),\\[1mm]
     \rm{(b)}\;\;\chi_{2n+1}\;\;\rm{and}\;\;\chi_{2n+2}\;\;
     \mbox{\rm have exactly}\;\;2n+2\;\;\mbox{\rm zeros on}\;\;[0,T),\\[1mm]
     \rm{(c)}\;\;\vartheta_{2n}\;\;\mbox{\rm and}\;\;\vartheta_{2n+1}\;\;
     \mbox{\rm have exactly}\;\;2n+1\;\;\mbox{\rm zeros on}\;\;[0,T).
     \end{array}
     $$
     The intervals $(\lambda_0, \mu_0), (\mu_1, \lambda_1),\ldots$ are called
     intervals of stability and the intervals
     $(-\infty, \lambda_0), (\mu_0, \mu_1), (\lambda_1, \lambda_2),\ldots$
     are called intervals of instability.

     \vspace{2ex}
     We use now (\ref{1.7}) and (\ref{1.8}) to rewrite operators
     $L_1$, $L_2$ in more
     appropriate form. From the expression for $r(x)$ from (\ref{1.7})
     and the relations
     between elliptic functions $sn(x)$, $cn(x)$ and $dn(x)$, we obtain
       $$L_1=\alpha^{2}[ -\partial_{y}^{2}+6k^{2} sn^{2}(y)-4-k^2] $$
     where $y=\alpha x$.

     It is well-known that the first five eigenvalues of
     $\Lambda_1=-\partial_{y}^{2}+6k^{2}sn^{2}(y, k)$,
     with periodic boundary conditions on $[0, 4K(k)]$, where
     $K(k)$ is the complete elliptic integral of the first kind, are
     simple. These eigenvalues and corresponding eigenfunctions are:
      $$\begin{array}{ll}
         \nu_{0}=2+2k^2-2\sqrt{1-k^2+k^4},
         & \phi_{0}(y)=1-(1+k^2-\sqrt{1-k^{2}
         +k^{4}})sn^{2}(y, k),\\[1mm]
         \nu_{1}=1+k^{2}, & \phi_{1}(y)=cn(y, k)dn(y, k)
         =sn'(y, k),\\[1mm]
         \nu_{2}=1+4k^{2}, & \phi_{2}(y)=sn(y, k)dn(y, k)
         =-cn'(y, k),\\[1mm]
         \nu_{3}=4+k^{2}, & \phi_{3}(y)=sn(y, k)cn(y, k)
         =-k^{-2}dn'(y, k),\\[1mm]
         \nu_{4}=2+2k^{2}+2\sqrt{1-k^{2}+k^{4}},
         & \phi_{4}(y)=1-(1+k^{2}+\sqrt{1-k^{2}
         +k^{4}})sn^{2}(y, k).
        \end{array}
      $$
      It follows that the first three eigenvalues of the operator
      $L_1$, equipped with periodic boundary condition on $[0,2K(k)]$
      (that is, in the case of left and right family),
      are simple and $\lambda_0=\alpha^2(\nu_0-\nu_3)<0, \;
      \lambda_1=\alpha^2(\nu_3-\nu_3)=0, \;
      \lambda_{2}=\alpha^2(\nu_4-\nu_3)>0$.
The corresponding eigenfunctions are
$\psi_0=\phi_0(\alpha x), \psi_1=r'(x), \psi_2=\phi_4(\alpha x)$.

\vspace{2ex}
Similarly, for the operator $L_2$ we have
$$L_2=\alpha^2[-\partial_y^2+2k^2sn^2(y, k)-k^2]$$
in the case of left and right family. The spectrum of
 $\Lambda_2=-\partial_y^2+2k^{2}sn^{2}(y, k)$ is formed
 by bands $[k^{2}, 1]\cup [1+k^{2}, +\infty)$. The
 first three eigenvalues and the corresponding eigenfunctions with
 periodic boundary conditions on $[0, 4K(k)]$ are simple and
       $$\begin{array}{ll}
          \epsilon_0=k^2, & \theta_0(y)=dn(y, k),\\[1mm]
          \epsilon_1=1, & \theta_1(y)=cn(y, k),\\[1mm]
          \epsilon_2=1+k^2, & \theta_2(y)=sn(y, k).
        \end{array}
      $$
>From (\ref{2.3}) it follows that zero is an eigenvalue of  $L_2$
and it is the first eigenvalue in the case of left and right family,
with corresponding eigenfunction $r(x)$.

    \vspace{2ex}
     {\bf 3. The estimate for $M_{2}$}
%%%%%%%%%%%%%%%%%%%%%%%%%%%%%%%%%%%%%%%%%%%%%%%%%%%%%%%%%%%%%%%%%%%%%%%%%%

   Below, we will denote by $\langle f,g\rangle=\int_0^T f(x)g(x)dx$
   and by $||f||$ the scalar product and the norm in $L^2[0,T]$.
   In the formulas that follow, we take $r=r(\bar{x})$ with an argument
   $\bar{x}=x-\xi+(a+3\omega)t$. From the previous section, we know that
   when considered in $[0,T]$, the operator $L_2$ has an eigenfunction
   $r$ corresponding to zero eigenvalue and the rest of the spectrum
   is contained in $(\alpha^2,\infty)$.

     The derivative of $d_{q}^{2}(u, \varphi)$ with respect to $\eta$
     at the point where the minimum is attained is equal to zero.
     Together with (\ref{2.3}), this yields
       \begin{equation}\label{3.7}
         \begin{array}{ll}
     0&=-i\omega\int_{0}^{T}{[ e^{i\omega \eta}\varphi_{x}
   \overline{h}_{x}- e^{-i\omega \eta}\overline{\varphi}_{x}h_{x}
   +q( e^{i\omega \eta}\varphi \overline{h}
   - e^{-i\omega \eta}\overline{\varphi}h)]}dx\\[2mm]
 &=2\omega Im\int_0^T(-\varphi_{xx}+q\varphi)e^{i\omega\eta}\overline{h}dx\\[2mm]
 &=2\omega Im\int_0^T((q+\omega^2+a+2r^2)r-2i\omega r')(h_1-ih_2)dx\\[2mm]
 &=-2\omega\int_0^T [(q+\omega^2+a+2r^2)rh_{2}+2\omega r' h_1]dx.
         \end{array}
       \end{equation}
We set $h_2=\beta r(\bar{x})+\theta$,  $\;\int_0^T\theta rdx=0$.
Substituting in (\ref{3.7}), we obtain
      $$0=\beta ||r||^2\left(q+\omega^2+a+\frac{2||r^2||^2}{||r||^2}\right)
      +2\int_0^T (\theta r^3+\omega r' h_1) dx$$
 Using that $\frac{2||r^2||^2}{||r||^2} \geq -a$ (see estimate A of the Appendix),
    we obtain the estimate
      $$\begin{array}{ll}
        |\beta|\,||r||& \displaystyle \leq 2 \frac{\left|
        \int_0^T(\theta r^3+\omega r' h_1)dx\right|}
        {(q+\omega^2)||r||}\\[5mm]
        &\displaystyle \leq \frac{2||r^3||\cdot ||\theta||+2|\omega|\,
        ||r'||\cdot  ||h_1||}{(q+\omega^2) ||r||}\\[3mm]
        & \leq m_0(||\theta||+||h_{1}||),
       \end{array}
     $$
     where $m_0=2m_1(a,\omega)/(q+\omega^2)$ and
     $$m_1(a,\omega)=\max\limits_{c\in[-\frac14a^2,0]}\left(
     \frac{||r^3||}{||r||}, \frac{|\omega|\,||r'||}{||r||},
     \frac{3||r^2r'||}{||r'||},\frac{|\omega|\,||ar+2r^3||}{||r'||}\right)$$
     (the third and fourth item are included for later use). It is obvious
     that the first three fractions are bounded. For the last one, see
     estimate D in the Appendix. We will use below that for $a$ and $\omega$
     fixed, $m_0\rightarrow 0$ when $q\rightarrow \infty$. Further,
       $$||h_{2}||\leq |\beta|\,||r||+||\theta||\leq m_0
       (||\theta||+||h_{1}||)+||\theta||=(m_0+1)||\theta||+m_0||h_1||.$$
     Hence, we obtain
     \begin{equation}\label{3.8}
       ||\theta||^{2}\geq \frac{||h_2||^2}{2(m_0+1)^2}
       -\left( {\frac{m_0}{m_0+1}}\right)^2||h_1||^2.
       \end{equation}
   Since $L_{2}r=0$ and $\langle \theta, r\rangle=0$, then from the spectral
   properties of the operator $L_2$, it follows
     $$
       M_2=\langle L_2 h_2, h_2\rangle =\langle L_2\theta, \theta\rangle\geq
       \alpha^2\langle \theta, \theta\rangle \geq -\frac{a}{2}||\theta||^2.
       $$
     From here and (\ref{3.8}), one obtains
       \begin{equation}\label{3.9}
           M_2 \geq \frac{|a|}{4(m_0+1)^2}||h_2||^2
           -\frac{|a|m_0^2}{2(m_0+1)^2} ||h_1||^2.
       \end{equation}

    \vspace{2ex}

%%%%%%%%%%%%%%%%%%%%%%%%%%%%%%%%%%%%%%%%%%%%%%%%%%%%%%%%%%%%%%%%%%%%%%%%%%
   {\bf 4. The estimate for $M_{1}$}

 First of all, let us note that the operator $L_1$ equipped with
 periodic boundary conditions in $[0,T]$ has the following spectral data:
\begin{equation}\label{spl1}
\begin{array}{ll}
\lambda_0=a-2\sqrt{a^2+3c},\quad & \psi_0=6r^2+3a-\lambda_0,\\
\lambda_1=0 & \psi_1=r',\\
\lambda_2=a+2\sqrt{a^2+3c},\quad & \psi_2=6r^2+3a-\lambda_2,
\end{array}\end{equation}
and the rest of the spectrum is contained in $(\lambda_2,\infty)$.

   We set
     \begin{equation}\label{3.10}
       h_1=\gamma_1\psi_0(\bar{x})+\gamma_2r'(\bar{x})
       +\theta_1, \; \; r(\bar{x})=\nu \psi_0(\bar{x})+\psi,
     \end{equation}
   where
   \begin{equation}\label{3.10.5}
   \langle \theta_{1}, \psi_{0}\rangle= \langle \theta_{1}, r'\rangle=
   \langle \psi, \psi_{0}\rangle= \langle \psi_0,r'\rangle=
   \langle \psi, r'\rangle=0
   \end{equation}
   and $\gamma_1$, $\gamma_2$ and $\nu$ are some constants.
   By (\ref{3.10.5}), we have
     $$M_{1}(h_{1})=\langle L_1h_1, h_1\rangle =\gamma_1^2\lambda_0
    \langle \psi_0,\psi_0\rangle+\langle L_1\theta_1, \theta_1\rangle.$$
    Therefore, from spectral properties of the operator $L_1$ it follows
 \begin{equation}\label{3.11}
 M_1(h_1)\geq \gamma_1^2\lambda_0||\psi_0||^2+\lambda_2||\theta_1||^2.
 \end{equation}
  The fundamental difficulty in the estimate of $M_1$ is the appearance
  of the negative term $\gamma_1^2 \lambda_0 ||\psi_0||^2$. Below, we
  are going to estimate it. From the condition
    $$P(u)=\int_0^T{|h+e^{i\omega \eta}\varphi(x-\xi,t)|^2}dx=P(\varphi)$$
   we obtain
     $$||h||^2=2Re\int_0^T e^{i\omega \eta}\varphi(x-\xi,t)\overline{h} dx
     =-2\int_0^T r h_1 dx.$$
  Then using (\ref{3.10}), we have
      $$-\frac12||h||^2=\nu \gamma_1||\psi_0||^2
      +\int_0^T \psi\theta_1 dx$$
      and therefore
   \begin{equation}\label{3.12}
   \gamma_1^2||\psi_0||^2=\frac{1}{\nu^2 ||\psi_0||^2}\left(\frac12||h||^2
        +\int_0^T \psi \theta_1 dx\right)^2.
        \end{equation}
 From (\ref{3.12}), we obtain
      \begin{equation}\label{3.13}
        \gamma_1^2||\psi_0||^2\leq \frac{1}{\nu^{2}
        ||\psi_0||^2}\left({\frac{1+d}{4}}||h||^{4}
        +{\frac{d+1}{d}}||\psi||^2||\theta_{1}||^2\right),
     \end{equation}
  where $d$ is a positive constant which will be fixed later.
  Below, we will denote by $C_m$, $D_m$ positive constants,
  depending only on $d$ but not on the system parameters $a, c, \omega$.
  Using (\ref{3.12}) and (\ref{3.11}), we derive the inequality
    \begin{equation}\label{3.14}
     \begin{array}{ll}
      M_1&\geq \left(\lambda_2+\lambda_0(1+\frac{1}{d})
      {\frac{||\psi||^2}{\nu^2||\psi_0||^2}}\right)||\theta_{1}||^{2}
      +{\frac{\lambda_0(1+d)}{4\nu^2||\psi_0||^2}}||h||^4 \\[2mm]
      &\geq C_1\lambda_2||\theta_1||^2-D_1|a|^\frac12||h||^4
     \end{array}
   \end{equation}
(see the estimates in point C of the Appendix).

We denote $\vartheta=h_1-\gamma_2r'(\bar{x})=\gamma_1\psi_0(\bar{x})+\theta_1$.
Then from (\ref{3.10.5}), (\ref{3.13}) and the inequalities
$\lambda_2\leq\frac13|\lambda_0|\leq |a|$, we have
    $$\begin{array}{rl}
    ||\vartheta||^2=\gamma_1^2||\psi_0||^2+||\theta_1||^2&
 \leq \left(1+\frac{(d+1)||\psi||^2} {d\nu^2||\psi_0||^2}\right)||\theta_1||^2
       +\frac{1+d}{4\nu^2||\psi_0||^2}||h||^4\\[2mm]
    &\leq C_2||\theta_1||^2+ D_2|a|^{-\frac12}||h||^4.\end{array}$$
    Then
      $$||\theta_1||^2\geq \frac{||\vartheta||^2}{C_2}
      -\frac{D_2||h||^4}{C_2|a|^\frac12}$$
      and hence, by (\ref{3.14}) and $\lambda_2\leq |a|$,
      \begin{equation}\label{3.15}
        \begin{array}{rl}
           M_1&\displaystyle\geq \frac{C_1\lambda_2}{C_2}||\vartheta||^2
            -\frac{C_1D_2+C_2D_1}{C_2}|a|^\frac12||h||^4\\[3mm]
            &= C_3\lambda_2||\vartheta||^2-D_3|a|^\frac12||h||^{4}.
        \end{array}
      \end{equation}

      After differentiating (\ref{3.2}) with respect to $\xi$, we obtain
        $$\begin{array}{ll}
  0&=2Re\int_0^T{e^{i\omega \eta}(\varphi_{xx}\overline{h}_x
  +q\varphi_x\overline{h})}dx
 =2Re\int_0^T{e^{i\omega \eta}\left[\varphi_{t}
 +(6|\varphi|^{2}+q)\varphi_x\right]\overline{h}}dx\\[2mm]
   &=2Re\int_0^T{(h_1-ih_2)[i\omega (3a+\omega^2+6r^2+q)r
   +(a+3\omega^2+6r^2+q)r']}dx\\[2mm]
  &=2\int_0^T{[(a+3\omega^2+6r^2+q)r'h_1+\omega(3a+\omega^2 +6r^2+q)rh_2]}dx.
           \end{array}
         $$
       From (\ref{3.7}), we have
 $$\int_0^T{qrh_2}dx=-\int_0^T[2\omega r'h_1+(a+\omega^2+2r^2)rh_2]dx$$
         and replacing in the above equality, we obtain
         $$\int_0^T{[(a+\omega^2+6r^2+q)r'h_1+\omega (2a+4r^2)rh_2]}dx=0.$$
       Substituting $h_1=\gamma_2r'(\bar{x})+\vartheta$ in the above
       equality and using the orthogonality condition
       $\langle r',\vartheta\rangle=\langle r', \gamma_1\psi_0
       +\theta_1\rangle=0$, we obtain
       $$\gamma_2||r'||^2\left(a+\omega^2+q+\frac{6||rr'||^2}
     {||r'||^2}\right)+2\int_0^T{[\omega (a+2r^2)rh_2+3r^2r'\vartheta]}dx=0.$$
         Using that ${\frac{6||rr'||^2}{||r'||^2}}\geq -a$ (see Appendix),
         we further have
           $$\begin{array}{ll}
        |\gamma_2|\,||r'||& \displaystyle\leq {\frac{2\left| \int_0^T
        {[\omega (a+2r^2)rh_2+3r^2r'\vartheta]}dx\right| }
        {(\omega^2+q) ||r'||}}\\[5mm]
   &\displaystyle \leq 2{\frac{|\omega|\,||ar+2r^3||\cdot||h_2||
      +3||r^2r'||\cdot ||\vartheta||}{(\omega^2+q)||r'||}}\\[4mm]
             &\leq m_0(||\vartheta||+||h_2||).
            \end{array}
          $$
 Hence
       $$||h_1||\leq|\gamma_2|\,||r'||
        +||\vartheta||\leq (m_0+1)||\vartheta||+m_0||h_2||,$$
        which yields
  $$||\vartheta||^{2}\geq \frac{||h_1||^2}{2(m_0+1)^2}
  -\left(\frac{m_0}{m_0+1}\right)^2||h_2||^2.$$
        Replacing in (\ref{3.15}), we finally obtain
          \begin{equation}\label{3.16}
   M_{1}\geq \frac{C_3\lambda_2}{2(m_0+1)^2}||h_1||^2
   -\frac{C_3\lambda_2m_0^2}{(m_0+1)^2}||h_2||^2-D_3|a|^\frac12|h||^4.
            \end{equation}

       \vspace{2ex}
        {\bf 5. The estimate for $\Delta M$}
%%%%%%%%%%%%%%%%%%%%%%%%%%%%%%%%%%%%%%%%%%%%%%%%%%%%%%%%%%%%%%%%%%%%%%%%%%%%%

        From (\ref{3.9}) and (\ref{3.16}), we have
    $$ M_1+M_2 \geq \frac{C_3\lambda_2-|a|m_0^2}{2(m_0+1)^2} ||h_1||^2
  +\frac{|a|-4C_3\lambda_2m_0^2}{4(m_0+1)^2}||h_2||^2-D_3|a|^\frac12||h||^4.$$
  We now fix $q$ so that $|a|m_0^2\leq\frac12C_3\lambda_2$ and assuming that
  $C_3\leq\frac12$ (which is no loss of generality), one has also
  $4C_3\lambda_2m_0^2\leq\frac12|a|$. Therefore we obtain
   $$M_1+M_2\geq C_4\lambda_2(||h_1||^2+||h_{2}||^2)-D_3|a|^\frac12||h||^4=
   C_4\lambda_2||h||^2-D_3|a|^\frac12||h||^4$$
   where $C_4$ and $D_3$ are absolute constants independent on the parameters
   of the system.

  On the other hand, estimating directly $I_{2}$ from below (for this purpose
  we use its initial formula), we have
       $$\begin{array}{ll}
       I_2&\geq ||h_{x}||^{2}+\int_{0}^{T}{(\omega^{2}-a-4r^{2})|h|^{2}}dx
       -2|\omega|\int_{0}^{T}{|h|\cdot |h_{x}|}dx
       -2\int_{0}^{T}{r^{2}|h|^{2}}dx\\[2mm]
       &\geq ||h_{x}||^{2}+(\omega^{2}-a+4a)||h||^{2}-2\omega^{2}||h||^{2}
       -{\frac{1}{2}}||h_{x}||^{2}+2a||h||^{2}\\[2mm]
       &={\frac{1}{2}}||h_{x}||^{2}-(\omega^{2}-5a)||h||^{2}.
       \end{array}
       $$
       Similarly, $|I_{3}|\leq \max(4|a|^\frac12|h|+|h|^2)||h||^2$.
       Let $0<m<\frac12$. We have
         $$\begin{array}{ll}
             \Delta M & =2mI_2+(1-2m)(M_1+M_2)+I_3\\[1mm]
             &\geq m||h_{x}||^2-2m(\omega^2-5a)||h||^2
             +(1-2m)(C_4\lambda_2||h||^2
             -D_3|a|^\frac12||h||^4)\\[1mm]
             &-\max (4|a|^\frac12|h|+|h|^2)||h||^2\\[1mm]
             &=m||h_{x}||^2+\left[-2m(\omega^2-5a)+
             (1-2m)C_4\lambda_2\right]||h||^2\\[1mm]
             &-[\max(4|a|^\frac12|h|+|h|^2)+(1-2m)D_3|a|^\frac12||h||^2]||h||^2.
             \end{array}
        $$
        We choose $m$, so that
          $$2qm=(1-2m)C_4\lambda_2-2m(\omega^2-5a),\;\;\mbox{\rm i.e.}\;\;
       2m=\frac{C_4\lambda_2}{q+\omega^2+5|a|+C_4\lambda_2}<1.$$
       From the inequality
         $$|h|^2\leq {\frac{1}{T}}\int_0^T{|h|^2}dx +2\left(\int_0^T{|h|^2}dx
         \int_0^T{|h_{x}|^2}dx \right)^\frac12$$
       we obtain
         $$|h|^2\leq {\frac{1}{T}}\int_0^T{|h|^2}dx+\sqrt{q}\int_0^T{|h|^2}dx
         +\frac{1}{\sqrt{q}}\int_0^T{|h_{x}|^2}dx.$$
         Hence for sufficiently large $q$, we obtain
       $$\max |h(x,t)|^2\leq {\frac{2}{\sqrt{q}}}d_{q}^{2}(u, \varphi )$$
    and moreover $||h||^2\leq q^{-1}d_{q}^{2}(u, \varphi )$. Consequently we
    can choose $\delta_{0}>0$, such that for $d_{q}(u, \varphi )<\delta_{0}$,
  we will have
  $[\max(4|a|^\frac12|h|+|h|^2)+(1-2m)D_3|a|^\frac12]||h||^2\leq qm$.

  Finally, we obtain that if $d_q(u, \varphi )<\delta_0$,
  then $\Delta M\geq md_q^2(u, \varphi)$.  Proposition \ref{p31} is
  completely proved.  $\Box$

       \vspace{2ex}
        {\bf 6.  Proof of Theorem \ref{t31}}
%%%%%%%%%%%%%%%%%%%%%%%%%%%%%%%%%%%%%%%%%%%%%%%%%%%%%%%%%%%%%%%%%%%%%%%%%%%%%

 We split the proof of our main result
 into two steps. We begin with the special case
      $P(u)=P(\varphi)$. Assume that $m,q, \delta_{0}$
     have been selected according to Proposition \ref{p31}. Since
     $\Delta M$ does not depend on $t, t\in [0, \infty)$, there
     exists a constant $l$ such that $\Delta M \leq ld^{2}(u,
     \varphi)|_{t=0}$. Below, we shall assume without loss of
     generality that $l\geq 1, q\geq 1$.

     Let
       $$\varepsilon >0, \; \; \delta = \min \left( \left(
       {\frac{m}{lq}}\right){\frac{\delta_{0}}{2}}, \left(
       {\frac{m}{l}}\right)^{1/2}\varepsilon \right) $$
     and $d(u, \varphi)|_{t=0}<\delta$. Then
       $$d_{q}(u, \varphi)\leq q^{1/2}d(u,
       \varphi)|_{t=0}<{\frac{\delta_{0}}{2}} $$
       and Lemma \ref{l31}  yields that there exists a
       $t_{0}>0$ such that $d_{q}(u, \varphi)<\delta_{0}$ if $t\in [0,
       t_{0})$. Then, by virtue of Proposition \ref{p31} we have
         $$\Delta M \geq md^{2}_{q}(u, \varphi), \; \; t\in [0,
         t_{0}).$$
         Let $t_{max}$ be the largest value such that
          $$\Delta M \geq md^{2}_{q}(u, \varphi), \; \; t\in [0,
         t_{max}).$$
         We assume that $t_{max}<\infty.$ Then, for $t\in [0,
         t_{max}]$ we have
         $$d^{2}_{q}(u, \varphi)\leq {\frac{\Delta M}{m}}\leq
         {\frac{l}{m}}d^{2}(u,
         \varphi)|_{t=0}<{\frac{l}{m}}\delta^{2}\leq
         {\frac{\delta^{2}_{0}}{4}}.$$
         Applying once again Lemma \ref{l31}, we obtain that there
         exists $t_{1}>t_{max}$ such that
           $$d_{q}(u, \varphi)<\delta_{0}, \; \; t\in [0,
           t_{1}).$$
           By virtue of the proposition, this contradicts the
           assumption $t_{max}<\infty$. Consequently,
           $t_{max}=\infty$,
           $$\Delta M \geq md^{2}_{q}(u, \varphi)\geq md^{2}(u,
           \varphi), \; \; t\in [0, \infty).$$
           Therefore,
             $$d^{2}(u, \varphi)\leq {\frac{\Delta M}{m}}\leq
             {\frac{l}{m}}\delta^{2}<\varepsilon^{2}, \; \; t\in
             [0, \infty), $$
             which proves the theorem in the special case.

             Now we proceed to remove the restriction
     $P(u)=||u||^{2}=||\varphi||^{2}=P(\varphi)$. We have (see (\ref{EK}))
     $||\varphi||=(2r_0E(k))^{1/2}$, where $r_0$ is given by (\ref{1.7}).
     Below, we are going to apply a perturbation argument, freezing for
     a while the period $T$ and the parameters $a,c$ in (\ref{newton}).
     We claim there are respective parameter values $a^*, c^*$, and corresponding
     $\varphi^*$, $r^*$,  $r_0^*$, $k^*$, see (\ref{1.3}), (\ref{newton}) and (\ref{1.7}),
     such that $\varphi^*$ has a period $T$ in $x$ and moreover,
     $2r_0^*E(k^*)=||u||^2$. By (\ref{1.7}), we obtain the equations
     \begin{equation}\label{ift}
     \begin{array}{l}
     \displaystyle \frac{2K(k^*)}{r_0^*}-T=0,\\
     \displaystyle 2r_0^*E(k^*)-||u||^2=0.
     \end{array}
     \end{equation}
If (\ref{ift}) has a solution $k^*=k^*(T, ||u||)$,  $r_0^*=r_0^*(T, ||u||)$,
then the parameter values we need are given by
$$a^*=({k^*}^2-2){r_0^*}^2,\quad c^*=({k^*}^2-1){r_0^*}^4.$$
Moreover, one has $||\varphi^*||=||u||$ and we could use the restricted
result we established above.
As $k=k^*(T, ||\varphi||)$,  $r_0=r_0^*(T, ||\varphi||)$, it remains to
apply the implicit function theorem to (\ref{ift}). Since the corresponding
functional determinant reads
$$\left|\begin{array}{cc}\frac{2K'(k^*)}{r_0^*} & -\frac{2K(k^*)}{{r_0^*}^2} \\
   2r_0^*E'(k^*)  &    2E(k^*)\end{array}\right|=\frac{4}{r_0^*}(KE)'
   =\frac{4}{r_0^*}({\textstyle\frac12}\pi+KK')>0$$
   (by Legendre's identity), the existence of $a^*$ and $c^*$ with the
   needed  properties is established.

%%%%%%%%%%%%%%%%%%%%%%%%%%%%%%%%%%%%%%%%%%%%%%%%%%%%%%%%%%%%%%%%%%%%%%%%%%%%%
By (\ref{ift}) and our assumption, we have
                \begin{equation}\label{4.20}
\frac{K(k^*)}{r_0^*}=\frac{K(k)}{r_0}=\frac{T}{2}.
                \end{equation}
Next, choosing $\eta=2(a^*-a)t$, $\xi=(a-a^*)t$, by (\ref{1.3}) and
(\ref{3.1}) one easily obtains the inequality
              $$d^2(\varphi^*, \varphi)\leq
              (1+\omega^2)||r^*-r||^2+||{r^*}'-r'||^2.$$
Denote for a while
$\Phi(\rho)=\rho\, dn(z\rho;k(\rho))
=\rho\, dn(y;k)$ where $k=k(\rho)$ is dertermined from
$K(k)=\frac12\rho T$. Then using (\ref{1.7}), we have
$r^*-r=\Phi(r_0^*)-\Phi(r_0)=(r_0^*-r_0)\Phi'(\rho)$ with some appropriate
$\rho$. Moreover,
$$\Phi'(\rho)=dn(y;k)+\rho\left[z\frac{\partial dn}{\partial y} (y;k)
+\frac{T}{2K'(k)}\frac{\partial dn}{\partial k}(y;k)\right]$$
satisfies $|\Phi'(\rho)|\leq C_0$ with a constant $C_0$ independent
on the values with $*$ accent.  Hence,
$|r^*-r|\leq C_0|r_0^*-r_0|$. Similarly,
$|{r^*}'-r'|\leq C_1|r_0^*-r_0|$. All this, together with (\ref{4.20}) yields

              \begin{equation}\label{4.21}
d(\varphi^*,\varphi)\leq C  |r_0^*-r_0|=\frac{2C}{T}|K(k^*)-K(k)|=
\frac{2C}{T}|K'(\kappa)||k^*-k|.
               \end{equation}

              From the inequalities
   $$\left|\, ||\varphi^*||-||\varphi||\,\right|=\left|\,
   ||u||-||\varphi||\,\right|\leq d(u,\varphi)|_{t=0}<\delta$$
    it  follows that
                 $$-\left( 2r_0E(k) \right)^{-1/2}\delta <
                 (||\varphi||)^{-1}||\varphi^*||-1<\left( 2r_0E(k)
                 \right)^{-1/2}\delta $$
                 and, therefore,
      $1-\delta_1<{\frac{r_0^*E(k^*)|}{r_0E(k)}}<1+\delta_1$,
 i.e. $|r_0^*E(k^*)-r_0E(k)|<r_0E(k)\delta_1$,
  where we have denoted $\delta_1=(1+(2r_0E(k))^{-1/2}\delta)^2-1$.

                 On the other hand, we have (using (\ref{4.20}) again)
                   \begin{equation}\label{4.22}
|r_0^*E(k^*)-r_0E(k)|=\frac{2}{T}|K(k^*)E(k^*)-K(k)E(k)|=
\frac{2}{T}|(KE)'(\kappa)||k^*-k|\geq C_2|k^*-k|,
                     \end{equation}
with  appropriate $C_2>0$ independent on the values bearing $*$ accent.
In particular, one has $|k^*-k|\leq C_3r_0E(k)\delta_1$.
 Thus combining (\ref{4.21}) and (\ref{4.22}), we get
  $$d(u, \varphi^*)|_{t=0}\leq d(u,
                     \varphi)|_{t=0}+d(\varphi,
                     \varphi^*)|_{t=0}<\delta+Cr_0E(k)\delta_1=\delta_0.$$

Let $\varepsilon >0$. We select $\delta$ (and together, $\delta_0$ and
$\delta_1$) sufficiently small and apply the part of the theorem which has
already been proved to conclude:
  $$d(u, \varphi^*)|_{t=0}<\delta_0\Rightarrow
                     d(u, \varphi^*)<{\frac{\varepsilon}{2}}, \;
                     \; t\in [0, \infty).$$
Then, choosing an appropriate $\delta>0$, we obtain that
                        $$d(u, \varphi)\leq d(u,
                     \varphi^*)+d(\varphi,
                     \varphi^*)<
   {\frac{\varepsilon}{2}}+Cr_0E(k)\delta_1<\varepsilon$$
                     for all $t\in [0, \infty)$. The theorem is
                     completely proved. $\Box$

\vspace{2ex}
  \section{Appendix}
%%%%%%%%%%%%%%%%%%%%%%%%%%%%%%%%%%%%%%%%%%%%%%%%%%%%%%%%%%%%%%%%%%%%%%%%%%

For $n\in\Z$ and $c\in(-\frac14a^2,0)$, consider the
line integrals $I_n(c)$ and their  derivatives $I'_n(c)$ given by
\begin{equation}\label{in}
I_n(c)=\oint_{H=c}x^nydx,\qquad I_n'(c)=\oint_{H=c}\frac{x^ndx}{2y}
\end{equation}
where one can assume for definiteness that the integration is along the
right oval contained in the level set $\{H=c\}$.
These integrals would be useful because
\begin{equation}\label{red}
\int_0^Tr^n(t)dt=2\int_0^{\frac12T}r^n(t)dt
=2\int_{r_1}^{r_0}\frac{x^ndx}{\sqrt{c-ax^2-x^4}}
=\oint_{H=c}\frac{x^ndx}{y}=2I_n'(c).
\end{equation}
(we applied a change of the variable $r(t)=x$ in the integral
and used equation (\ref{newton})).
The properties of $I_n$ are well known, see e.g. \cite{HIK}
for a recent treatment. Below, we list some facts we are going to use.

\vspace{2ex}
\noindent
{\bf Lemma.}  (i)  {\it The following identity holds:
$$(n+6)I_{n+3}+(n+3)aI_{n+1}-ncI_{n-1}=0,\quad n\in\Z$$
which implies}
\begin{equation}\label{recur}
\textstyle
I_3'=-\frac12a I'_1,\quad
I_4'=\frac13cI_0'-\frac23aI_2',\quad
I_6'=-\frac{4}{15}acI_0'+(\frac{8}{15}a^2+\frac35c)I_2'.
\end{equation}

\vspace{1ex}
\noindent
(ii) {\it The integrals $I_0$ and $I_2$ satisfy the system}
$$\begin{array}{l}
4cI'_0-2aI'_2=3I_0,\\
-2acI'_0+(12c+4a^2)I'_2=15I_2.\end{array}$$

\vspace{1ex}
\noindent
(iii) {\it The ratio $R(c)=I_2'(c)/I_0'(c)$ satisfies the Riccati equation
and related system
\begin{equation}\label{ricc}
(8c^2+2a^2c)R'(c)=ac+4cR(c)-aR^2(c),\qquad
\begin{array}{l}\dot{c}=8c^2+2a^2c,\\
\dot{R}=ac+4cR-aR^2,\end{array}\end{equation}
which imply estimates}
\begin{equation}\label{est}
\frac{2c}{a}\leq R(c)\leq\frac{c}{2a}-\frac{3a}{8}.
\end{equation}

\vspace{2ex}
\noindent
The equations in (i)--(iii) are derived in a standard way, see \cite{HIK}
for more details. The estimates (\ref{est}) follow from the fact that,
in the $(c,R)$-plane,
the graph of $R(c)$ coincides with the concave separatrix trajectory of
the system (\ref{ricc}) contained in the triangle with vertices $(0,0)$,
$(-\frac14a^2,-\frac12a)$ and $(0,-\frac38a)$ and connecting the first
two of them.

After this preparation, we turn to prove the estimates we used in the
preceding sections.

\vspace{2ex}
\noindent
{\bf A. The estimate for } $A=\frac{2||r^2||^2}{||r||^2}$.
By (\ref{red}), (\ref{recur}) and the first inequality in (\ref{est}),
we have
$$A=\frac{2\int_0^Tr^4dt}{\int_0^Tr^2dt}=\frac{2I_4'}{I_2'}=
\frac{2cI_0'-4aI_2'}{3I_2'}=\frac{2c}{3}\frac{1}{R}-\frac{4a}{3}\geq-a.$$

\vspace{2ex}
\noindent
{\bf B. The estimate for} $B=\frac{6||rr'||^2}{||r'||^2}$.
By (\ref{newton}), we have as above
$$B=\frac{6\int_0^Tr^2(c-ar^2-r^4)dt}{\int_0^T(c-ar^2-r^4)dt}
=\frac{6(cI_2'-aI_4'-I_6')}{cI_0'-aI_2'-I_4'}$$
$$=\frac65\frac{(2a^2+6c)I'_2-acI'_0}{2cI'_0-aI'_2}
=\frac65\frac{(2a^2+6c)R-ac}{2c-aR}\geq-\frac{12}{5}a.$$
To obtain the last inequality, we used that $4c+a^2\geq 0$
and the second estimate in (\ref{est}).

\vspace{2ex}
\noindent
{\bf C. The estimate for}
$C=\lambda_2+\lambda_0(1+\frac{1}{d})\frac{||\psi||^2}{\nu^2||\psi_0||^2}$.
By (\ref{3.10}) and (\ref{3.10.5}) we have
$$||\psi||^2=||r||^2-\nu^2||\psi_0||^2,\;\;\mbox{\rm where}\;\;
\nu=\frac{\langle r,\psi_0\rangle}{||\psi_0||^2}.$$
Therefore
$$C=\lambda_2+\lambda_0\left(1+\frac{1}{d}\right)
\left(\frac{||r||^2||\psi_0||^2}{\langle r,\psi_0\rangle^2}-1\right).$$
Next,
$$\langle r,\psi_0\rangle=\int_0^T[6r^3+(3a-\lambda_0)r]dt
=12I_3'+(6a-2\lambda_0)I'_1=-2\lambda_0I'_1,$$
$$\begin{array}{rl}
||r||^2||\psi_0||^2 &=\int_0^Tr^2dt\int_0^T(6r^2+3a-\lambda_0)^2dt\\[4mm]
&=4I_2'[36I'_4+12(3a-\lambda_0)I'_2+(3a-\lambda_0)^2I_0']\\[2mm]
&=4I_2'[(12c+(3a-\lambda_0)^2)I'_0+(12a-12\lambda_0)I'_2].
\end{array} $$
By (\ref{1.7}), we have
\begin{equation}\label{EK}
I_2'(c)=\frac12\int_0^Tr^2dt=r_0\int_0^{K(k)}dn^2(t)dt=r_0E(k).
\end{equation}
Making use of the identity $E(k)=\frac12\pi F(\frac12,-\frac12,1,k^2)$
where $F$ is the Gauss hypergeometric function, we obtain an appropriate
expansion to estimate $E$ from above
$$
E(k)=\frac{\pi}{2}\left(1-\frac{k^2}{4}-\frac{3k^4}{64}-\frac{5k^6}{512}
-\ldots,\right), \quad
E^2(k)\leq\frac{\pi^2}{4}\left(1-\frac{k^2}{2}-\frac{k^4}{32}\right)$$
with all removed terms negative. As $I_1'=\frac12\pi$, by (\ref{1.7})
this implies
$$I_2'^2\leq - I_1'^2\frac{a^2+20ar_0^2+4r_0^4}{32r_0^2}.$$
Together with $I_0'I_2'\geq I_1'^2$, this yields
$$\begin{array}{rl}\displaystyle
\displaystyle \frac{||r||^2||\psi_0||^2}{\langle r,\psi_0\rangle^2}-1
&\displaystyle \leq\frac{1}{\lambda_0^2}\left[12c+(3a-\lambda_0)^2
+\frac{3}{8r_0^2}(\lambda_0-a)(a^2+20ar_0^2+4r_0^4)\right]-1\\[3mm]
&=\displaystyle\frac{\lambda_2}{\lambda_0}\left(\frac{\lambda_2-a}{8r_0^2}-1\right)
\leq\frac{\lambda_2}{\lambda_0}\left(\frac{\sqrt3}{8}-1\right)\end{array}$$
where the equality is obtained by direct calculations.
Therefore,
$$C\geq \lambda_2\left(-\frac{1}{d}+\frac{d+1}{d}\frac{\sqrt{3}}{8}\right)
=C_1\lambda_2$$
with $C_1>0$ an absolute constant when $d\geq 4$ is fixed.

As a by-product of our calculations, we easily obtain also the estimate
$$\frac{\lambda_0}{\nu^2||\psi_0||^2}=
\frac{\lambda_0||\psi_0||^2}{\langle r,\psi_0\rangle^2}
\geq \frac{\lambda_2(\frac{\sqrt3}{8}-1)+\lambda_0}{||r||^2}
\geq -D_1|a|^\frac12.$$

\vspace{2ex}
\noindent
{\bf D. The estimate for } $D=\frac{||ar+2r^3||}{||r'||}$.
Making use of statements (i) and (ii) of the Lemma, we have
$$D^2=\frac{\int_0^T(a^2r^2+4ar^4+4r^6)dt}{\int_0^T(c-ar^2-r^4)dt}=
\frac{a^2I_2'+4aI_4'+4I_6'}{cI_0'-aI_2'-I_4'}$$
$$=\frac{4acI_0'+(7a^2+36c)I_2'}{5(2cI_0'-aI_2')}
=\frac{aI_0+6I_2}{I_0}\leq -5a.$$

%%%%%%%%%%%%%%%%%%%%%%%%%%%%%%%%%%%%%%%%%%%%%%%%%%%%%%%%%%%%%%%%%%%%%%%%%%


\begin{thebibliography}{99}

 \bibitem{AlBoHe}J. Albert,  J.L. Bona, D. Henry,  Sufficient
conditions for stability of solitary-wave solutions of model
equations for waves, {\it Physica D} {\bf 24} (1987), 343--366.


\bibitem{An2} J. Angulo,  Nonlinear stability of periodic travelling
wave solutions to the Schr\"odinger and the modified Korteweg-de
Vries equations, {\it J. Differential Equations} {\bf 235} (2007),
1--30.

\bibitem{ABS} J. Angulo, J.L. Bona, M. Scialom, Stability of
cnoidal waves, {\it Adv. Differential Equations} {\bf 11} (2006),
1321--1374.

\bibitem{AnNa} J. Angulo, F. Natali, Positivity properties of the
Fourier transform and the stability of periodic travelling-wave
solutions, {\it SIAM, J. Math. Anal.} {\bf 40} (2008), 1123--1151.

 \bibitem{Be1} T.B. Benjamin,  The stability of solitary waves,
 {\it Proc. R. Soc. London} Ser. A {\bf 328} (1972), 153--183.

 \bibitem{BKPS} B. Birnir, C. Kenig, G. Ponce, N. Svenstedt, On
 the ill-posedness of the IVP for the generalized Korteweg-de
 Vries and nonlinear Schr\"odinger equation, {\it J. London Math.
 Soc.} {\bf 53} (1996), 551--559.

 \bibitem{BPS}B. Birnir, G. Ponce, N. Svenstedt, The local
 ill-posedness of the modified KdV equation, {\it Ann. Inst. H.
 Poincar\'{e} (Anal. Non Lin\'{e}aire)} {\bf 13} (1996), 529--535.

 \bibitem{Bo} J.L. Bona, On the stability theory of solitary waves,
 {\it Proc. R. Soc. London} Ser. A {\bf 344} (1975), 363--374.

 \bibitem{BGT} N. Burq, P. Gerard, N. Tzvetkov, An instability
 property of the nonlinear Schr\"{o}dinger equation on $S^{d}$,
 {\it Math. Res. Lett.} {\bf 9} (2002), 323--335.


 \bibitem{GSS} M. Grillakis, J. Shatah, W. Strauss,  Stability
 of solitary waves in the presence of symmetry I, {\it J. Funct.
 Anal.} {\bf 74} (1987), 160--197.

\bibitem{HIK} S. Hakkaev, I.D. Iliev, K. Kirchev,
 Stability of periodic travelling shallow-water waves determined by Newton's
 equation, {\it J. Phys. A: Math. Theor.} {\bf 41} (2008), 31 pp.

 \bibitem{Ha} M. H\v{a}r\v{a}gu\c{s}, Stability of periodic
 waves for the generalized BBM equation, {\it Rev. Roumanie Math.
 Pure Appl.} {\bf 53} (2008), 445--463.

 \bibitem{IK} I.D.  Iliev, K.P. Kirchev,  Stability and instability
 of solitary waves for one-dimensional Schr\"odinger equations,
{\it  Differential Integral Equations} {\bf 6} (1993), 685--703.

\bibitem{Ka1} T. Kato, Quasilinear equations of evolution, with
applications to partial differential equations, 25--70, in: {\it Spectral Theory
and Differential Equations}, Proc. Symp. Dundee, 1974, Lecture
Notes in Math. {\bf 448}, Springer-Verlag.

\bibitem{Ka2} T. Kato, On the Korteweg-de Vries equation, {\it
Manuscr. Math.} {\bf 28} (1979), 89--99.

\bibitem{KPV1} C. Kenig, G. Ponce, L. Vega, On the ill-posedness
of some canonical dispersive equations, {\it Duke Math. J.} {\bf
106} (2001), 617--633.

 \bibitem{MaWi} W. Magnus, S. Winkler, {\it Hill's Equation},
   Interscience, Tracts in Pure and Appl. Math. {\bf 20},  Wiley, NY,
   1976.

\bibitem{W1} M. Weinstein, Lyapunov stability of ground states of
nonlinear dispersive evolution equations, {\it Comm. Pure Appl.
Math.} {\bf 39} (1986), 51--68.

\bibitem{W2} M. Weinstein, Existence and dynamic stability of
solitary-wave solutions of equations arising in long wave
propagation, {\it Commun. Partial Diff. Eqns.} {\bf 12} (1987),
1133

 \bibitem{ZIK} E.P. Zhidkov, I.D. Iliev, K.P. Kirchev,
 Stability of a solution of the form of a solitary wave for a
 nonlinear complex modified Korteweg-de Vries equation,
 {\it Sib. Mat. Zh.} {\bf 26}(1985), no. 6, 39--47 [in Russian].



 \end{thebibliography}
 \end{document}